\documentclass[shortnames]{tise_style}

\usepackage{graphics}
\usepackage{amssymb}
\usepackage{amsmath}
\usepackage{mathrsfs}
\usepackage{graphicx}
\usepackage{setspace}
\usepackage{tikz}
\usepackage{subfigure}

\author{Amelia McNamara\\Smith College}
\title{On the State of Computing in Statistics Education: Tools for Learning and for Doing}
\Abstract{Here is a lovely abstract}
\Plainauthor{\@author}
\Volume{VV}
\Year{YYYY}
\Month{MMMMMM}
\Issue{II}
\Submitdate{yyyy-mm-dd}
\Acceptdate{yyyy-mm-dd}
\Address{
}

 \Seriesname{Issue}
  \Hypersubject{Journal of Statistical Software}
  \Plaintitle{\@title}
  \Shorttitle{\@title}
  \Plainkeywords{\@Keywords}

  \Hyperauthor{\@Plainauthor}
  \Keywords{---!!!---at least one keyword is required---!!!---}
  \Footername{Affiliation}
  \Firstdate{\textit{Submitted:} \@Submitdate}
  \Seconddate{\textit{Accepted:} \@Acceptdate}

\begin{document}

\section{INTRODUCTION}

When Rolf Biehler wrote his 1997 paper, ``Software for Learning and for Doing Statistics,'' some educators were already using computers in their introductory statistics classes~\citep{Bie1997}. Biehler's paper laid out suggestions for the conscious improvement of software for statistics, and much of his vision has been realized. But, computers have improved tremendously since 1997, and statistics and data have changed along with them. 

It has become normative to use computers in statistics courses-- the 2005 Guidelines for Assessment and Instruction in Statistics Education (GAISE) college report suggested that students ``Use real data'' and ``Use technology for developing conceptual understanding and analyzing data''~\citep{AliCobCuf2005}.  In 2010, Deb Nolan and Duncan Temple Lang argued students should ``Compute with data in the practice of statistics''~\citep{NolLan2010}. Indeed, much of the hype around `data science' seems to be centering around computational statistics. Donoho's vision of `Greater Data Science' is almost entirely dependent on computation~\citep{Don2015}. The 2016 GAISE college report updated their recommendations to ``Integrate real data with a context and purpose'' and ``Use technology to explore concepts and analyze data''~\citep{CarEve2016}. So, it is clear that we need to be engaging students in statistical computing. But how? 

To begin, we should consider how computers are currently being used in statistics classes. Once we have a solid foundation of the current state of computation in statistics education, we can begin to dream about the future. 

 \subsection{The Gap Between Tools for Learning and Doing}\label{gap}
When thinking about statistical computation in education, we are considering the transition from a novice to a more experienced student. For the purposes of this paper, we focus primarily on introductory statistics courses at the college level. However, it should be noted that a novice is a novice. Whether a student is in high school, college, or graduate school, if they have not seen statistics or statistical computation before, they will experience a similar learning curve. 

Tools that get used in statistics courses can be broken into two categories: tools for learning statistics, and tools for doing statistics~\citep{Bag2013, McN2015}. Both types of tools are used to introduce novices to statistical computing, and of course, they have their advantages and disadvantages. Tools designed for learning statistics are generally not good for actually performing data analysis, and tools for professionals tend to be hard to learn.

This distinction between types of tools has been longstanding, and is often quite divisive. In his history of statistical computing, Jan De Leeuw explicitly states he is only concerned with ``statistical software'' and not ``software for statistics''~\citep{DeL2009} while~\cite{Bie1997} focuses only on novices' ability to grasp a tool with minimal instruction. Whatever your perspective, it is clear there is gap between these two types of tools. The tension between learning and doing has been discussed in the past~\citep{Bag2013, McN2015}, but research into the transition between the two states is still nascent. 

\subsection{Bridging the Gap}
The argument is often made that the two types of tools should be kept separate~\citep{Bie1997, Kon2007, Fri2008}. In particular, Konold says tools for learning statistics should not be stripped down versions of professional tools for doing statistics. Instead, they should be developed with a bottom-up perspective, thinking about what features novices need to build their understandings~\citep{Kon2007}. 

However, we take the perspective that the gap must be either bridged or closed. In considering how to close the gap, we will keep in mind how tools can build novices' understanding from the ground up, but we will aim to end the siloing of tools for learning and for doing statistics. There are many approaches that could be used to do this. For example, curricular resources making explicit reference to a prior tool and couching tasks in terminology from the previous system might make the transition between technology easier. Likewise, providing some sort of `ramping up' as users reach the end of the abilities of the learning tool and `ramping down' at the beginning of the tool for doing could make the gap feel less abrupt. 

We could also imagine a new type of tool, bridging from a supportive tool for learning to an expressive tool for doing. This tool could be used by novices and experts alike. However, there are few models of software in other domains that manage this balance.  

Considering the barriers (both technological and philosophical), an easier way to imagine bridging the gap is to consider tools for learning and for doing statistics `reaching across' the gap, aiming to take good qualities from their counterpart across the divide.

\subsection{Evaluating currently-existing tools}
\label{attributes}
In order to begin thinking about how to develop new tools or improve existing ones, we must consider the strengths and weaknesses of the currently-existing landscape of tools. \cite{McN2016c} provides a critical framework through which to evaluate tools, laying out a set of 10 attributes necessary for a modern statistical computing tool. These attributes are 

\begin{enumerate}
\item Accessibility 
\item Easy entry for novice users 
\item Data as a first-order persistent object 
\item Support for a cycle of exploratory and confirmatory analysis 
\item Flexible plot creation 
\item Support for randomization throughout 
\item Interactivity at every level 
\item Inherent documentation 
\item Simple support for narrative, publishing, and reproducibility 
\item Flexibility to build extensions 
\end{enumerate}

As we consider tools, these attributes will be invoked. 

\subsection{Types of Currently-Existing Tools}
In this paper, we will attempt to consider the current landscape of currently-available statistical tools, from the prosaic (Excel) to the bespoke (Wrangler, Lyra). Keeping in mind the gap between tools for learning and tools for doing statistics, and the attributes listed in Section \ref{attributes}, we will attempt to assess the field of existing technology for learning and doing statistics.

In statistics education, a distinction is often made between route-type and landscape-type tools for statistical computing~\citep{Bak2002}. Route-type tools drive the trajectory of learning, while landscape-type tools are less directive, and allow users to explore the field of possibilities themselves~\citep{Bak2002}. We will consider both route-type tools (e.g. applets, which only allow for one concept to be explored) and landscape-type tools (e.g., standalone educational software and statistical programming tools).

Drawing on the categories of technology proposed by the GAISE reports~\citep{AliCobCuf2005, CarEve2016} and the distinctions offered by Biehler~\citep{Bie1997}, we will consider the following types of technology:
\begin{itemize}
\item Graphing calculators
\item Spreadsheets
\item Applets and microworlds
\item Standalone educational software
\item Statistical programming tools
\item Tools for reproducible research
\item Bespoke tools
\end{itemize}

\section{GRAPHING CALCULATORS}
In statistics, there are some educators who believe the mathematical underpinnings of statistics are sufficient to provide novices with an intuitive understanding of the discipline. Because of this, they do not find it imperative that statistics education be accompanied by computation. One example to consider is the Advanced Placement Statistics course and associated exam. The AP Statistics teacher guide states ``students are expected to use technological tools throughout the course,'' but goes on to say technological tools in this context are primarily graphing calculators~\citep{CB2008}. Instead of building in computer work, the guide suggests exposing students to computer output so they can learn to interpret it.  

In this paradigm, students learn basic concepts about sampling, distributions, and variability, and work through formulas by hand. They use calculators to assist with their arithmetic calculations. 

Calculators should not be considered appropriate tools for statistical computation. In particular, they make it impossible to work with real data, to say nothing of providing data `as a first-order persistent object'~\citep{McN2016c}. Additionally, the analysis that is produced is not reproducible, and the `computation' does not help students develop a deeper understanding of the underlying concepts. 

\section{SPREADSHEETS}\label{spreadsheets}
Spreadsheet software like Excel are probably the most commonly used tools to do data analysis by people across a broad swath of use-cases~\citep{Bry2016}. While the Microsoft Office Suite that contains Excel can be expensive, the free availability of spreadsheet tools like Google Spreadsheets and the Open Office analogue of Excel, Sheets, means spreadsheets can be considered to satisfy the attribute of accessibility~\citep{McN2016c}. 

However, spreadsheets lack the functionality to be a true tool for statistical programming. They typically allow for only limited scripting, which means their capabilities are limited to those built in by the development company. The locked-in nature of the functionality means they are only able to provide a limited number of analysis and visualization methods, and cannot be flexible enough to allow for true creativity. They fail at providing `flexible plot creation,' or the `flexibility to build extensions'~\citep{McN2016c}. 

Beyond this, several high profile cases of academic paper retraction have been based on internal errors within Excel~\citep{HerAsh2013}. Because the underlying code is closed-source, Excel does not allow users to view how methods are implemented, which means it is very difficult for an individual to assess the validity of the internal code. Some dedicated researchers have tested Excel's statistical validity over every software version Microsoft has released. Not only is every version flawed, but even with specific attention shed on the problem, Microsoft often either fails to repair the problem, or makes a change to another flawed version~\citep{McCHei2008, Mel2014}. 

Additionally, spreadsheets do not privilege data `first-order, persistent object'~\citep{McN2016c}.  Once a data file is open, modification or deletion of data values is just a click away. In this paradigm, the sanctity of data is not preserved, and original data can be lost forever.  In contrast, most statistical tools discourage direct manipulation of original data. In tools used by practitioners to do statistical analysis (e.g., \proglang{R}, SAS software), data is an almost sacred object, and users are given only a copy of the data to work with.

Data does not have structural integrity in a spreadsheet. Data values sit next to blocks of text and plots produced by data cover up data cells. Everything is included on one canvas. These pieces may be linked together, but there is no explicit visual connection. In a true statistical tool, results from the analysis are separated from the data from which they were derived, and any data cleaning tasks performed in these tools can be easily documented. 

This leads to the largest challenge with spreadsheets: they do not support reproducibility~\citep{McN2016c}. Data journalists have historically done analysis using tools like Excel~\citep{PlaCoo2015}. Journalists must be careful about the analysis they publish, as it must be verified like any other `source' they might interview. Spreadsheets do not offer any inherent documentation. As a result, journalists developed their own reproducibility documentation, often called a `data diary.' The data diary typically takes the form of a document written in parallel with the analysis that describes all the steps taken. This supplementary document is done separately, either by hand or in word processing software like Microsoft Word.  There are ways to improve this process~\citep{WilBry2016}, but it is still inherently precarious. 

Because each stage of analysis in a spreadsheet is done by clicking and dragging, there is no way to fully record all the actions taken. One of the central tenets of reproducibility is it should be possible to perform the same analysis on slightly different data (e.g., from a different year). Spreadsheets do not make this possible, so they are not effective tools for data analysis.  

However, spreadsheets have some clear advantages. The first is their great market saturation, as mentioned above. Almost everyone has access to some sort of spreadsheet program, and many levels of schooling offer spreadsheet training. 

The second is that because output and plots are created by linking cells of the data to attributes in the analysis, the final product is reactive. When you change a value in the spreadsheet, all the associated graphs update automatically. Computer scientist Alan Kay believes computer operating systems should essentially be spreadsheets, by which he means an operating system should be a reactive programming environment that can be built up into responsive tools to perform a wide variety of tasks~\citep{Kay1984}. In this paradigm, objects can be linked together in a dependent structure, and whenever an input is changed, all the downstream elements are updated accordingly. 

Again, because so many people have access to spreadsheet tools, the interactive product can be easily shared with others. Although the dependent structures are not made visible in the spreadsheet, readers of the product can get a sense of the connections by playing with the data. 

Of course, the reactive possibilities in spreadsheets can also lead to unintended consequences. In a study of spreadsheets used by Enron, researchers found 24\% of spreadsheets with a formula included an error \citep{HerMur2015}. This is likely because while spreadsheets allow for reactive linking of cells, they do not visualize the reactive connections, and it can be easy to double a formula or include unintended cells. The reactive programming environment \pkg{Shiny} showcases some of the capabilities of this paradigm in a more reproducible data analysis environment~\citep{ChaChe2015}. 

Overall, while spreadsheets are widely accessible and have reactive capabilities, they should not be considered true tools for statistical programming. This is because they do not privilege data, making it easy to accidentally (or intentionally) manipulate the original values, and contain errors in their closed-source code.

\section{APPLETS AND MICROWORLDS}\label{applets}

\subsection{Applets}
\label{applets}
Statistics applets are typically hosted on the web, and are programmed to illustrate one concept through the use of a specialized interactive web tool. They are highly accessible because they are hosted online, and they are typically free to use. Biehler would consider these `microworlds'-- tools that allow instructors to curate a limited list of functions for students to use.

Some of the best applets were designed by statistics educators Alan Rossman and Beth Chance~\citep{ChaRos2006}. One of their applets allows students to discover randomization by working through a scenario about randomly assigning babies at the hospital. The applet asks the question, \emph{if we randomly assign four babies to four homes, how often do they end up in the home to which they belong?} The user can watch the randomization happen once, as a stork flies across the screen to deliver the babies to their color-coded homes, and then accelerate the illustration to see what the distribution would look like if one tried the same experiment many times. Students can use checkboxes to turn the animation on or off, or to see the theoretical probabilities. They can also try again with a different number of babies or a different number of trials. 

Another popular applet set called StatKey was developed by the Lock5 group to accompany their textbook~\citep{Lock5, LocLoc2014}. One StatKey applet allows students to create a bootstrap confidence interval for a mean. This applet does not include an animation like the stork featured in the Rossman-Chance example, but users can still specify how many samples they want to take, stepping through one sample at a time, or accelerating the process by clicking the ``generate 1000 samples'' button. 

Applets can be useful for students to learn distinct concepts like randomization, but they can also be frustrating when students want to do things just outside the scope of the applet. The Rossman and Chance applets include many concepts, but very few of them let users import their own data. The StatKey applets do allow users to edit the example data sets or upload entirely new data, but they are necessarily limited to what they were programmed to do. In other words, they fail on the attributes of the `flexibility to build extensions' and `support for a cycle of exploratory and confirmatory analysis'~\citep{McN2016c}

StatCrunch is another popular tool used by educators. It was developed by Webster West, and it combines features from standalone software packages like TinkerPlots and Fathom alongside spreadsheet-like functionality and `microworld'-style applets. StatCrunch is inexpensive and available through the web browser, so it is accessible. It was initially developed in the late 1990s as a \proglang{Java} applet~\citep{WesWu2004}, but has since been reworked into a modern web tool (likely using \proglang{JavaScript}) distributed by Pearson Education. While StatCrunch does collect a lot of the best features of the tools it amalgamates, it also accumulates many of the negatives. For example, the lack of data sanctity mentioned in the spreadsheets section is certainly true here, as is the messy canvas associated with both spreadsheets and software like TinkerPlots and Fathom.

\subsection{Interactive Data Visualizations}\label{interact}
Interactive data visualizations are gaining popularity on the web. The New York Times produces some especially salient examples. Instead of static graphics, their visualizations allow readers to manipulate representations of data themselves. For example, the Times has produced graphics allowing readers to balance the federal budget, predict which way states will vote in the presidential election, or assess whether they would save more money by buying or renting their housing~\citep{CarEri2010,ElectoralMap2012, BosCar2014}. 

These data visualizations are essentially applets or microworlds. They allow a user to learn about one particular facet of a dataset the author has made available. This scripted quality is actually valued in data visualization, because visualizations should provide some context and storytelling for the data, rather than simply leaving users to explore~\citep{Cai2013}. But the script can also be limiting.

Data visualizations can serve much the same purpose as applets (helping users understand one particular concept) but have the same drawbacks (inflexibility). 

Some interactive visualizations have been pushing the boundaries on this inflexibility, allowing readers to critique the creation process or algorithmic decisions. One notable example is the IEEE Spectrum rating of programming languages. The article provides a default ranking, but it allows readers to create a custom ranking by adjusting the weights of all the data inputs~\citep{CasDiaRom2014}. It is possible to imagine a future where all journalistic products based on data are accompanied by this type of auditable representation of the process used to create them.

\subsection{Shiny and manipulate}\label{shiny}
A recent addition to this ecosystem are the \proglang{R} packages \pkg{Shiny} and \pkg{manipulate}, which would likely be termed ``meta-tools'' by Biehler, enabling teachers to ``adapt and modify material and software for their students'' \citep{Bie1997}. (\proglang{R} and its package system are discussed in more detail in Section \ref{R}.) 

\pkg{Shiny} enables \proglang{R} programmers to create interactive visualizations for the web~\citep{ChaChe2015}. Authoring \pkg{Shiny} apps is a task for more expert \proglang{R} users, but the resulting applets are similar to those described in Section \ref{applets}, so they can be useful teaching tools for novices to play with. The applets are reactive, which puts some burden on the programmer but means that they are highly interactive. 

\pkg{Shiny} has enabled \proglang{R} programmers to build interactive tools that have gained viral success, such as the dialect map published by the New York Times that eventually received more views than any article in the history of the paper~\citep{KatAnd2013, Leo2014}. There are also many examples of applets developed specifically for education~\citep{Cec2014}. 

\pkg{Shiny} supports interface features like sliders, radio buttons, check boxes, and text input. Typically, though, the resulting visualizations are themselves static. The user cannot zoom into them naturally in the way they would with a other interactive web graphics. Instead, the programmer would have to incorporate sliders for the x- and y-ranges, and the user would manipulate those to impact the zoom. In other words, \pkg{Shiny} apps suffer from many of the same drawbacks as more traditional applets, although they are easier for instructors to develop. 

A simpler \proglang{R} package with a similar idea is the \pkg{manipulate} package~\citep{All2014}. \pkg{manipulate} is easier for novices to use, although it still requires some knowledge of \proglang{R}. However, instead of producing a standalone interactive graphic, \pkg{manipulate} works within RStudio to produce interaction for the purposes of education.

\subsection{Reflections on applets and microworlds}
While applets can become frustrating for students because of their limited scope, the array of well-considered statistics education applets is a rich source of inspiration for projects trying to bridge the gap. Applets tend to be the most successful at satisfying the attribute of `interactive at every level,' although they are rarely flexible enough to build extensions~\citep{McN2016c}. If statistical programming tools could aim for some of the interactivity and animation of applets, they might be better understood by all users. \pkg{Shiny} also shows promise, because it makes it simpler for educators to create their own applets to illustrate new concepts. Ideally, it should be possible for anyone to create an interactive data product, not just those with have \proglang{R} or \proglang{JavaScript} skills.

\section{STANDALONE EDUCATIONAL SOFTWARE}\label{tinkerplots}
The field of standalone educational software is dominated by sibling software packages TinkerPlots and Fathom. Although computers were being used in introductory statistics classes before Biehler's 1997 paper, it seems clear there was a turning point after it was published. Fathom and TinkerPlots, designed by Cliff Konold and William Finzer, respectively, can both trace their origins to ``Software for Learning and for Doing Statistics'' and have realized much of Biehler's vision~\citep{Bie1997}. 

TinkerPlots and Fathom have been around for years, and are well-loved but becoming outdated. Two more recent developments in the field of standalone educational software are CODAP (a new project by Finzer and Konold) and iNZight. We'll talk about each of these in turn. 

\subsection{TinkerPlots and Fathom}
The authors of Fathom and TinkerPlots wanted to design tools relevant to the way students think. The two have similar functionality, although slightly different intended users. TinkerPlots is described as being appropriate for students from 4th grade up to university, and Fathom is directed at the secondary school and introductory college levels. Because Fathom is intended for slightly older users, it includes more features than TinkerPlots does.


Both TinkerPlots and Fathom are excellent tools for novices to use when learning statistics. They comply with nearly all the specifications outlined by \cite{Bie1997}, allowing for flexible plotting, providing a low threshold, and encouraging play and re-randomization. They allow students to jump right in, to perform exploratory data analysis and to move through a data analytic cycle (e.g., asking questions, trying to answer them, re-forming questions), and have been shown to enhance student understanding~\citep{WatDon2009}. 


Fathom was developed by William Finzer, based on principles from~\cite{Bie1997}, and intended to allow students play with statistical concepts in a more creative way. The design specs upon which Fathom is based include a focus on resampling, a belief there should be no modal dialog boxes, the location of controls outside the document proper, and animations to illustrate what is happening~\citep{Fin2002b}.  

TinkerPlots was designed by a team led by Clifford Konold, a psychologist focused on statistics education~\citep{KonMil2005}. TinkerPlots was built on Fathom's infrastructure, but designed for younger students. TinkerPlots was developed the same year as the initial GAISE report~\citep{FraKadMor2005}, and the connection between the cognitive tasks TinkerPlots makes possible and the A and B levels of the guidelines is clear. TinkerPlots includes probability modeling, but no standard statistical models (e.g. linear regression). Users can develop their own simulations and link components together to see how changing elements in one area will impact the outcome somewhere else. 

TinkerPlots and Fathom have a large market share when it comes to teaching introductory statistics in the K-12 context~\citep{Leh2007, GarBen2008, KonKaz2008, WatFitz2010, BieBenBak2013, Fin2013, Fit2013, MatRee2013}, at the introductory college level~\citep{Ben2000, GarCha2002, EveZie2008} and in training for teachers~\citep{Rub2002, Bie2003, GouPec2004, HamRub2004,  RubHamKon2006, Hal2008, PfaBen2011}. Past their design principles, both tools were popular for their reasonable pricing strategy, which made it possible for schools to afford licenses. They are accessible in terms of pricing and compatibility with many platforms, but they may not be useable by students with disabilities~\citep{McN2016c}. 


For educators who want to teach concepts like randomization and data-driven inference, the primary competitors at this level are applets. TinkerPlots and Fathom have a number of advantages over applets. Most importantly, TinkerPlots and Fathom allow students to use whatever data they want, rather than demonstrating data on one locked-in data set. The systems come with pre-loaded data sets, but it is easy to open other data and use it in the same way. 

%

Fathom was developed in 2002, and TinkerPlots in 2005. In the 10 years since their respective releases, statistical programming has moved forward in ways these packages have not. For example, while few would expect novices to be working with `big data' in the truest sense of the term, TinkerPlots can only deal with data up to a certain size. A trial using a dataset with 12,000 observations and 20 variables caused considerable slowing, while larger datasets caused the program to hang indefinitely. Fathom dealt with the same data much more easily, but still had a noticeable delay loading and manipulating the data. 

While both software packages allow for the inclusion of text in the workspace, there is no way to develop a data analysis narrative. They do not `support narrative, publishing, and reproducibility'~\citep{McN2016c}. The more free-form workspace can feel creative, but it makes it nearly impossible to reproduce analysis, even using an existing file. There is also no easy way to publish results from these programs. The proprietary file types (.tp for TinkerPlots and .ftm for Fathom) need the associated software in order to be run interactively, and the only way to produce something viewable without the application is to print the screen. 

Because the software is closed-source, neither TinkerPlots nor Fathom are extendable in any way. There is no `flexibility to build extensions.' What you see is what you get. This becomes particularly problematic when it comes to modern modeling techniques. For example, in the Introduction to Data Science class developed for high school students through the Mobilize grant, students use classification and regression trees, and perform k-means classification~\citep{GouJohn2015}. Those methods are not available in either software package, and cannot be added. In fact, TinkerPlots has no standard statistical models, which means it cannot be used for the full data analytic cycle. It is truly only a tool for learning. Fathom, which is designed for slightly older students, does provide limited modeling functionality in the form of simple linear regression and multiple regression. 

In the context of Clifford Konold's argument that tools for learning should be completely separate from tools for doing~\citep{Kon2007}, it makes sense there are limits to these tools. They were consciously designed to be separate. However, given the capabilities of modern computing, it should be possible to provide this ground-up entry while still supporting more extensibility. 

\subsection{CODAP}
Both William Finzer and Cliff Konold are involved with the  development of the Common Online Data Analysis Platform (CODAP). CODAP promises to be a more modern, web-based take on the principles that drove the development of Fathom and TinkerPlots~\citep{Con2016}. While CODAP is still in the initial stages of development, it shows promise. Several partnerships with data-generating groups have formed initial test cases for CODAP. One partner is the group Ocean Tracks, whose goal is to involve high school students in marine biology~\citep{Con2016}. Several other partnerships are in progress as well, and the CODAP team hopes to learn what features are most useful for a platform by generalizing from these specific examples.  Because CODAP is web-based, it will be more accessible than tools like Fathom and also make it easier to share and publish results of analysis. Only time will tell if it will support reproducibility and extensibility. 

\subsection{iNZight}
Another recent development in the category of standalone educational software is the microworld iNZight. Underlying iNZight is \proglang{R}, but unlike other graphical user interfaces (GUIs) for \proglang{R} (discussed in more detail in Section \ref{RGUIs}), the goal of iNZight is not for students to learn \proglang{R}. Instead, the designers of iNZight have used \proglang{R} as a target language for the development of a free, interactive statistical tool~\citep{WilEll2016}.  

iNZight is launched from within \proglang{R}, but launches its own GUI window. Unlike the \proglang{R} GUIs in Section \ref{RGUIs}, iNZight does not generate \proglang{R} code associated with the actions taken in the GUI. The tool has been designed to make it possible to do a variety of data analytic actions associated with statistics curriculum in New Zealand. 

While it is somewhat awkward to have to launch an application from another application that will not be explicitly used, otherwise iNZight delivers on being a lightweight, open-source, graphical approach to statistics. It is accessible, extensible, and interactive~\citep{McN2016c}. 

\subsection{Reflections on standalone educational software}
The field of standalone educational software for statistics has long been dominated by TinkerPlots and Fathom. These tools revolutionized the way statistics could be done by novices, and realized Rolf Biehler's 1997 vision. However, as computers and data have changed, they are becoming outdated. Luckily, both Finzer and Konold are involved with the Common Online Data Analysis Platform (CODAP) project, which has the potential to modernize the best aspects of the tools. Another promising development is iNZight, a microworld built on \proglang{R}, without exposing students to syntax. 

In the context of~\cite{McN2016c}, standalone educational software exemplifies many of the attributes. Because of their low price, these tools can be considered at least mostly accessible, although they do not offer accessibility support for disabled users. They certainly ease entry, and support exploratory and confirmatory analysis. They provide flexible plotting and make it simple to use randomization. They are interactive and visual. But, they are not appropriate tools to use to actually `do' data analysis. Data is not a persistent object in these tools, as cases can be easily changed with no record made. In the same vein, they do not support reproducible research, narrative, or publishing (to view a Fathom or TinkerPlots document, a reader must have the software installed on their computer). They are also closed-source and cannot be extended. 

Again, CODAP is aiming to solve some of these issues. Because it is open-source and on the web, documents will be accessible to anyone with internet access. However, educators choosing to use these tools should consider the tradeoffs between standalone educational software and professional tools.

\section{DATA DESK}
\label{datadesk}
Data Desk has been given its own section in this text because while it was considered to be standalone statistics education software by Rolf Biehler, it also is included in a history of statistical programming tools by Jan DeLeeuw~\citep{DeL2009}. It is the only program to garner this broad acceptance across a variety of use-cases.

 Data Desk was developed by Paul Velleman (a student of John Tukey) to facilitate Tukey's exploratory data analysis~\citep{Vel1989}.  It represents one of the first uses of linked visualization~\citep{Wills2008}. The first version of Data Desk was introduced in 1985, and although most of what persists from that version are screenshots, it is easy to see how groundbreaking it must have been at the time. When Biehler mentions Data Desk in his 1997 paper, development had been underway for more than 10 years~\citep{Bie1997}. 

In contrast, almost all other tools used for teaching statistics came after the 1997 paper.  Amazingly, Data Desk has gone through seven versions since 1985, and still exists today. The only other tool that might be able to claim such a long history is JMP (described  in Section \ref{S-tools}), although JMP did not have interactive graphics until 1991~\citep{BesMor1991}. 

The Data Desk interface was clearly an inspiration for TinkerPlots and Fathom, and features a palette of tools as well as menu bars. However, Data Desk provides much richer functionality than either TinkerPlots or Fathom, including linear and nonlinear models, cluster analysis, and principal component analysis. 

The drawbacks of Data Desk are slight, and similar to those of TinkerPlots and Fathom. After 30 years, the interface looks outdated, and it does not `support narrative, publishing, and reproducibility' because it does not promote the inclusion of text and only static versions of analysis can be shared~\citep{McN2016c}.  However, these drawbacks notwithstanding, it is a highly inspirational tool. 

\section{STATISTICAL PROGRAMMING TOOLS}
Over time, there has been a movement toward students as true `creators' of computational statistical work, which requires the use of a statistical programming tool. Deb Nolan and Duncan Temple Lang argue well for this in their paper, ``Computing in the statistics curriculum,'' where they suggest students should ``compute with data in the practice of statistics''~\citep{NolLan2010}.  They are promoting \proglang{R}, although their recommendations could be achieved using a different language. Many colleges and universities are modifying their statistics courses to fall in line with Nolan and Temple Lang's recommendations.

Statistical programming tools are those that can be used by statistical practitioners to do data analysis. Common examples include SAS software, SPSS software, Stata, \proglang{R}, \proglang{Python}, and \proglang{Julia}. The first three we will consider together, because these are enterprise software tools that can be prohibitively expensive (a major barrier to use in education). The last three (all programming languages) will be considered as open-source alternatives that are more `accessible'~\citep{McN2016c}.

\subsection{SAS software, Stata software, SPSS, and JMP}\label{S-tools}

Some commonly used tools for doing statistical analysis are SAS software, Stata software, SPSS, and JMP~\citep{SAS2015, Stata2015, SPSS2013, JMP2016}. Interestingly, although most users refer to these tools by shorthand names (e.g. `SAS') their official names include the word `software' (e.g. 'SAS software').  All three tools are standalone software, and all combine elements of graphical user interfaces with command-line tools. They are used in a variety of disciplinary contexts, so the argument for teaching them is `students will need to use this in the future.' They are often popular in industry, because they come with guarantees of validity and technical support, and they are designed for work with big data. However, they are closed-source, expensive, and can be unintuitive for novices.

SAS software and Stata software are perhaps the most similar of these three packages. Their interfaces are visually similar, and they have similar benefits and drawbacks. 

SAS software was first introduced in the 1970s~\citep{DeL2009}, when data was read in on punch cards. Because each card could only hold a small amount of information, the system necessarily had to be good at distributed computing. This functionality has scaled well over the years, and as a result SAS software is very powerful for working with big data. SAS software is often used in pharmaceutical and business applications because it comes with a guarantee of accuracy. SAS software has a command line interface (CLI) because of its long history, and the fact it was developed before personal computers were commonplace. For many years, it was difficult to re-use results from analysis as data, but the development of the SAS Output Delivery System (ODS) in the late 1990s changed that~\citep{BryMul2000}. The CLI and ability to use results as data supports reproducibility, which is discussed in more depth in Section \ref{repro-s}. 

The main drawback to SAS software is its price. The company makes the software free for educational use, both as desktop software and via the cloud, so students can access it via a web browser (which makes it more accessible). But, SAS software is hugely expensive for corporate use, in part because of the guarantee of accuracy and included support. Business pricing is not available on their website. Instead, users must submit a request for a quote.  

Stata software was first released in the 1980s, and initially only had a command line interface~\citep{DeL2009}. It is often the tool of choice for economists and, therefore, introductory statistics courses taught in economics departments. Since 2003, it has included a graphical user interface (GUI) in addition to the CLI~\citep{DeL2009}. The CLI can be used to create analyses can be re-run to get the same results, so it does support reproducible research. The tools that support reproducibility in Stata software are discussed in more depth in Section \ref{repro-s}. Stata's user interface is often thought of as more user-friendly than that of SAS. Software support is available in the system or via a phone number users can call to get personalized help. Stata software does provide the `flexibility to build extensions'~\citep{McN2016c}, and it has an archive of contributed code from users~\citep{DeL2009}. 

Like SAS software, the drawback to Stata software is its price. Although the company has reduced the price of almost all their products, they can still be cost-prohibitive for students and academic institutions. As of 2016, individual student pricing was \$125 for an annual license to the version that works with moderate-sized data or \$198 for a perpetual license. A single business license costs \$595 per year or \$1,195 for a perpetual license. The version of Stata software that works with large datasets (up to 10,998 variables) comes at an additional cost. The company does offer group discounts, but these are also expensive. Instructors using Stata software have complained that the licenses for educational use are often so limited their students must double up on computers to avoid getting kicked off the system. 

Similar to SAS software and Stata software, SPSS is another corporate tool for statistics. It is typically used by social scientists and is much more focused on a menu-driven interface, than SAS software and Stata software. SPSS does have a proprietary command-line syntax, but the syntax is harder for humans to parse and the code is generally only created by copying and pasting, versus users generating code themselves~\citep{ATS2013}. Although code can be saved, SPSS does not support reproducible research in the sense of literate programming or dynamic documents. SPSS is also very expensive-- \$1,170-\$7,820 for a 12-month individual license (pricing depending on features included), or \$69.99-\$89.99 for a one-year student license (plus \$4.99 download fee). 

The final product in this category is another offering from SAS Institute, called JMP~\citep{JMP2016}. JMP is often used in an educational setting, and provides a drag-and-drop, menu-driven graphical user interface to SAS software. Like Data Desk, discussed in Section \ref{datadesk}, JMP was originally designed in the 1980s and provides many features useful for novices, like interactive brushing and linking, generalizable data cleaning, and visual model support. The backbone of JMP is SAS software, so the analysis done in JMP can be considered to be reproducible, but JMP provides a simple visual interface. 

JMP provides many of the features of software for learning statistics with the power of a tool for really doing statistics. Like TinkerPlots and Fathom, while JMP does produce interactive graphics within an individual session, these interactive results cannot be exported. Instead, a work session can be printed or pasted into a document. The student version of JMP does not support exporting graphics, but individual licenses do. JMP provides a lot of inspiration for what an interactive statistical programming tool could look like, particularly one coupled with reproducible results. However, once again JMP is expensive (\$1,620 for an individual and \$14,900 for the professional version) but they offer academic discounts: \$49.95 for a 12-month license for undergraduate and graduate students.

Although SAS software, Stata software, and SPSS are commonly used in industry, none of them seem supportive of learners. They all provide only specific types of graphics (failing to provide `flexible plot creation'), and most work is done using menus and wizards, so they do not make clear what the tool is actually doing (no `inherent documentation')~\citep{McN2016c}. These tools creates `users' rather than `creators' of statistics. All three tools obscure the underlying computational processes and reduce statistical procedures to button clicks. They all provide some capability of extending the software with scripting, but they all suffer from a lack of transparency about how internal routines were coded. Finally, their pricing is prohibitive for many use-cases. JMP is the most inspirational of the group, providing graphical methods to interact with analysis as well as reproducible code in the SAS software language. 

\subsection{R}\label{R}
\proglang{R} is a programming language for statistical computing~\citep{RCore2016}. It is the tool of choice of academic statisticians, and has a growing market outside academia~\citep{Van2009}. Analysts at companies like Google routinely use \proglang{R} to perform exploratory data analysis and make models. 

\proglang{R} has several advantages over the other professional tools we have discussed. First, it is free. When members of the open source community use the word ``free'' they often distinguish between ``free as in speech'' and ``free as in beer.'' These phrases indicate the difference between software that costs no money (e.g., most Google products) and software that is completely unrestricted and available for anyone to modify and edit. \proglang{R} is free in both ways. It is also compatabile with computer accessibility features, making it, for example, useable by blind people~\citep{God2013}. Because of this, it is one of the only tools that can be considered to be fully `accessible'~\citep{McN2016c}. 

Like many programming languages, \proglang{R} has both a base language and additional libraries that extend its functionality, called packages. Most packages are hosted on a centralized server called the Comprehensive R Archive Network (CRAN)~\citep{CRAN2015}. CRAN makes it simple for users to install new packages. Because \proglang{R} has the statistical community invested in it, and because it is open-source and easy to modify, there are many additional packages for \proglang{R}.  As of this writing, CRAN hosts over 9,000 packages. Packages \pkg{mosaic}, \pkg{dplyr} and \pkg{ggplot2} are discussed in Section \ref{rsyntaxes}, and \pkg{Shiny} was already discussed in Section \ref{shiny}. 

Another great quality of \proglang{R} is that it makes it very difficult to modify original data (it privileges data as a `first-order, persistent object'). When working in \proglang{R}, a user is never interacting directly with the original data, rather a copy of the data that has been loaded into the work session. \proglang{R} keeps a history of all commands that have been used in a session, making it simpler to follow the trail of actions from the original data to a cleaned version and final analysis. \proglang{R} facilitates reproducible research, as discussed in Section \ref{knitr}. 

As with any tool, \proglang{R} has its shortcomings as well. The main drawback of \proglang{R} is its status as a programming language. Many of the other tools discussed here are graphical user interfaces (GUIs), while \proglang{R} is a language. Programming languages tend not to offer `easy entry' because they require users to provide syntactically correct function calls with appropriate arguments, and are not flexible about things like capitalization and punctuation. On top of this standard hurdle, \proglang{R} has an inconsistent syntax, which can make it particularly hard to master.

There have been efforts to simplify the coding aspects of \proglang{R} over the years. Some of these efforts are curricular, reducing the number of commands to which novices are exposed, or providing more consistent syntax ~\citep{Ver2005, KapSho2013}. Other efforts are GUIs like Deducer~\citep{Fel2012} and RCommander~\citep{Fox2004}. However, none of these efforts have truly solved the problem. 


\subsubsection{R syntaxes}\label{rsyntaxes}

One complex aspect of \proglang{R} is the multitude of syntaxes it supports. Where most programming languages would have one standard syntax, \proglang{R} has many. We will discuss three syntaxes that are commonly encountered, as well as how they fit into the attributes from~\cite{McN2016c}. 

Historically, \proglang{R} has used the `dollar sign syntax,' which uses the \verb#$# operator. For example, \\ \verb#diamonds$color# indicates the \verb#color# variable within the \verb#diamonds# dataset. This is often accompanied by subsetting using square brackets, as in \verb#diamonds[1, ]# which would pull the first row of the data or \verb#diamonds[ ,1]# which would pull the first column. This syntax does not provide easy entry or inherent documentation, because the symbols do not hold prior meaning. 

In educational settings, many teachers use the formula syntax, so named because it is most commonly found in functions performing modeling. Project MOSAIC and its associated \proglang{R} package, \pkg{mosaic}, promote the formula syntax and \pkg{mosaic} rewrites summary statistics functions to follow the convention~\citep{PruKap2015, PruKap2015b}. The formula syntax uses a \verb#~# operator, and instead of referring to variables within datasets, the user refers to the variables directly and then notes the dataset later. For example, \verb#tally(~color, data=diamonds)# counts the number of diamonds of each color in the dataset. By using the \pkg{mosaic} package, along with \pkg{lattice} graphics~\citep{Sar2008}, students can stay firmly within the formula-based syntax for an entire introductory college statistics course. At the high school level, the Mobilize project has also limited its scope to the formula syntax~\citep{GouJohn2015}. Limiting the scope of a course to the formula syntax seems to ease entry, and make it clearer to users what the code does. It is still difficult for novices to understand the leading \verb#~# in one-variable situations, but overall the formula syntax seems easier for users. 

One newer syntax (which can be mixed in with either paradigm mentioned above) is the pipe. This operator, \%\verb#>#\%, `pipes' data from one function into another~\citep{BacWic2014}. The pipe is most often used in conjunction with the so-called `tidyverse.' The tidyverse, so named because it works with `tidy' data~\citep{Wic2014}, includes the data ingestion package \pkg{readr}, data visualization package \pkg{ggplot2}, data manipulation packages \pkg{dplyr}, \pkg{lubridate}, \pkg{stringr}, and modeling package \pkg{broom}~\citep{WicFra2015, Wic2009, Wic2015, GroWic2011, Wic2016b, Rob2016}. Many of these packages were authored by Hadley Wickham, who has said he wants to build tools that allow users to easily express 90\% of what they want to be able to do, while only losing 10\% of the flexibility~\citep{Wic2014c}. Staying within the particular syntax of the tidyverse is essentially using a domain-specific language for data analysis Since the tidyverse is situated within the full-featured language of \proglang{R}, edge cases can be addressed with extensions as they are needed. 

An example of the tidyverse syntax with the pipe operator would be
\begin{verbatim}
diamonds %>%
    group_by(color) %>%
    tally()
\end{verbatim}
The pipe paradigm allows users to avoid writing dollar signs, so it is beginning to gain traction within the statistics education community. 

From these descriptions of the various syntaxes of \proglang{R}, it should be clear that they conform to the attributes from \cite{McN2016c} to various degrees. The formula and pipe syntaxes have the easiest entry, and better support the cycle of exploratory and confirmatory analysis. Plots from the \pkg{lattice} package, using the formula syntax, make it more difficult to do flexible plot creation than \pkg{base} plots associated with the dollar sign syntax or \pkg{ggplot2} plots.

\subsubsection{Graphical User Interfaces and Integrated Development Environments for R}
\label{RGUIs}
\proglang{R} work typically takes place at a command-line interface (CLI). In fact, \proglang{R} can be used directly at the command line or terminal. However, there are some Graphical User Interfaces (GUIs) and Integrated Development Environments (IDEs) for \proglang{R} that help support users in their work. 

In contrast to the CLI paradigm that characterizes much of programming, GUIs and IDEs reduce cognitive load on users by allow users to interact with computers by the use of menus and buttons (in the case of a GUI) or providing a source code editor with colored code, debugging support, code completion, and sometimes automated code refactoring (IDEs). Unlike standard source code editor (e.g. \proglang{vi} or Notepad++), IDEs provide additional support for programmers.   A common example of an language-specific IDE is Eclipse. Other IDEs are language agnostic, like Visual Studio Code by Microsoft, or Sublime Text.

The most common GUIs for \proglang{R} are R Commander and Deducer~\citep{Fox2004, Fel2012}, and the runaway winner in the IDE category is RStudio. 

Both R Commander and Deducer produce \proglang{R} code when graphical elements of the interface are manipulated, but do not move any further in encouraging users to transition from users to doers of statistics. They do not facilitate play (as Fathom does) or develop computational thinking (as learning \proglang{R} does). The connection between actions taken with the menus and the resulting code is implicit rather than explicit, and there is little reason for users to manipulate the code. 

R Commander was developed by John Fox as a way to use \proglang{R} in introductory statistics classes without students having to learn syntax~\citep{Fox2004}. It provides a limited set of possible tasks and provides a graphical user interface. While the interface is much more route-type than the landscape-type TinkerPlots and Fathom~\citep{Bak2002}, R Commander also makes it possible to do summary statistics, graphics, and simple models. 

Deducer is another GUI for \proglang{R} which, much like R Commander, provides access to some of \proglang{R}'s functionality through menus and wizards~\citep{Fel2012}. The plot menus produce beautiful \pkg{ggplot2} graphics~\citep{Wic2009}, which is ideal for users because they are inclined to feel pride for having created something appealing. However, the resulting \pkg{ggplot2} code printed into the console is too difficult for users to parse. Another detail is the automatically-created \proglang{R} code, which should ideally be non-threatening to encourage users to associate the actions they have taken in wizards with the resulting code. In Deducer, this auto-generated code is bright red. In many users' minds, red signifies `error' ~\citep{EllMai2007}, so users often initially think they have done something wrong. 

While neither R Commander or Deducer are ideal for teaching students computational thinking or facilitating play, they are freely available and work with a variety of systems. 

Much more useful for learners (though less graphical) is RStudio, an IDE for \proglang{R}~\citep{RStudio2014}. Like other IDEs, RStudio colors code to make it easier to parse, provides code completion, and makes it easier to debug. 

RStudio can be run as a desktop application for Mac, Windows, or Linux, but it is also available as a server install. With a server distribution, users go to a website, log in, and find their RStudio session just how they left it. The interface looks nearly identical to the desktop version, but contained within a browser window. All the data files and code are hosted on a central server, so students can do their work from any computer without having to worry about moving data from place to place. 

A server also allows instructors to manage package installations from a central location and provide quick bug fixes to all students at once. Because of the simple access and management, many colleges use RStudio servers for their students. In particular, Smith College, Mount Holyoke College, Duke University, and Macalester College all use this arrangement~\citep{BauCet2014}. At the high school level, the Mobilize project also used a server version to reduce startup friction for high school teachers and their students~\citep{GouJohn2015}. 

RStudio provides additional support features that improve on the standard \proglang{R} GUI. In particular, RStudio is a unified interface where windows cannot get `lost.' It also provides visual cues; to objects in the working environment, to installed packages, and to files in the working directory. It offers file management and comprehensive code history. The data preview functionality helps ease the transition from spreadsheet programs. And even in the most programming-oriented area, the Console, RStudio provides coding support features like tab completion and code hints, which increase the inherent documentation of \proglang{R}~\citep{McN2016c}. RStudio has been used successfully in many introductory college statistics classes \citep{BauCet2014, MulKid2014, PruHor2014, HorBau2014} and with high school teachers and students through the Mobilize Project~\citep{GouJohn2015}. However, even though it lowers the barrier to entry for \proglang{R}, RStudio still requires users to code, so there is a startup cost associated with using it.

\subsection{Python}\label{ipython}
\proglang{Python} is a general-purpose programming language that provides support to statistics through the packages \pkg{pandas}, \pkg{NumPy}, and \pkg{matplotlib}~\citep{McK2012}. It does not have the same statistical programming community that \proglang{R} does, but does have a large community considering reproducible research, discussed in more depth in Section \ref{jupyter}. 

Computer science education research has shown that \proglang{Python} is easier for novices to learn than \proglang{Java}, which has led many computer science departments to switch the language used in their introductory class to promote access~\citep{Guo2014, AlvDod2012, RanMil2006}. However, the language features that make \proglang{Python} easier to learn also apply to \proglang{R}-- because they are both interpreted languages with low startup costs. While \proglang{Python} is clearly advantageous in the introductory computer science context, it is not clear whether it is the appropriate tool for introductory statistics. 

\subsection{Julia}
Another programming language specifically designed for statistical computing is \proglang{Julia}. Unlike \proglang{R}, \proglang{Julia} is being built from programming language principles, so the authors hope to avoid many of the pitfalls \proglang{R} has run into over the years~\citep{BezEde2014}. 

Another principle \proglang{Julia} is based on is the idea that the language used by a \proglang{Julia} developer should be the same as the language used by a \proglang{Julia} user~\citep{BezEde2014}. In some scientific computing languages, the underlying code is written in another, faster language, like \proglang{FORTRAN} or \proglang{C++}. \proglang{R} falls into this category, as many of its faster routines are written in lower-level languages. This is akin to the `flexibility to build extensions' principle taken to the extreme~\citep{McN2016c}. 

The distinctions between \proglang{R} and \proglang{Julia} have not been fully fleshed-out yet. \proglang{Julia} is a new language, still under active development. Of course, this means that it does not have the user community that \proglang{R} has gathered over the course of 20+ years. However, \proglang{Julia} seems more aimed at computer scientists and programmers, rather than simply people with data problems looking to answer questions.

\subsection{Reflections on statistical programming tools}
Statistical programming tools are good for, simply, statistical programming. In this category we contrasted more GUI-driven tools like Stata software, SAS software, SPSS, and JMP, and programming languages \proglang{R}, \proglang{Python} and \proglang{Julia}. 

While SAS software, Stata software, SPSS and JMP are all GUI-driven, they all have underlying code which can be used to script analysis.  These tools are used by people in a variety of domain areas for solving statistical problems. Because of their prohibitive price, they tend to be used by corporations and in industry. Considering the 10 attributes from ~\cite{McN2016c}, they are not accessible, they often do not support easy entry, but they do privilege data as a first-order, persistent object. They support the cycle of exploratory and confirmatory analysis as well as any current tools do. While it can be challenging, they offer the possibility to do randomization, as well produce reproducible reports. Since they are scriptable, they offer the flexibility to build extensions. They are not good for interactivity (with the exception of JMP, the GUI to SAS software), nor inherent documentation. 

Programming languages like \proglang{R}, \proglang{Python} and \proglang{Julia} have many of the same pros and cons as the more graphical software packages, although they are more accessible. The most popular choice of language for statistics courses is \proglang{R}, which has the power of the statistics community driving package development. Because it is a programming language, \proglang{R} can be hard to learn, but efforts like Project MOSAIC have been working to make it more accessible to novices. RStudio smoothes out setup issues and provides a graphical view of data, as well as other friendly interface features. 
 \proglang{Python} has been shown to be novice-friendly in introductory computer science classes, but it doesn't have as much inherent support for statistical work. \proglang{Julia} is being written specifically for use in statistics, but it is so new it is hard to comment on. 

All these professional statistical programming tools have the power to work with large, arbitrary data and can produce reproducible code (unlike applets and standalone educational software). However, their high startup costs make them somewhat less than ideal for a learning context. 

\section{TOOLS FOR REPRODUCIBLE RESEARCH}
Reproducible research is a crucial element of statistics and statistics education. In the 2016 Guidelines for Assessment and Instruction in Statistics Education College Report, support for reproducible research is listed as one of the considerations for teachers when selecting tools~\citep{CarEve2016}. Previously, tools for reproducible research were intimately tied to the statistical package being used for analysis. However, most tools are now open to a variety of languages or software, so we can consider these tools separately from the target languages they grew out of.

There are many views of reproducibility. In this section, will be considering the narrowest view, that it should be possible to re-run the an analysis using the same data and code to get the same result again. It is said that `data wrangling' can take up to 80\% (or more!) of the time in data projects~\citep{KanHeePla2011}, so it is important that effort not be wasted. Ideally, it should be possible to run the analysis using a slightly-modified version of the data (for example, the next year of data collection) to get analogous results.  

Programming languages inherently support reproducible code. However, even with a script containing the analysis code or a history of all commands run, it is easy for code to become un-reproducible. Typically, this is because of human error-- the code gets separated from the analysis, so it is not clear which parts of the code correspond to which plots and output in the paper, or the code becomes outdated and is not human-readable enough to debug. To solve these issues, Donald Knuth proposed an idea of `literate programming' where all programs would be accompanied by surrounding narrative to explain to humans what they were doing~\citep{Knu1984}. 

Deborah Nolan and Duncan Temple Lang took this further, defining a dynamic document as one that is compiled and automatically includes the results of embedded code~\citep{NolTem2007}.  When we consider reproducible research in education, we want students to be creating dynamic documents. In the same paper, Nolan and Temple Lang define interactive documents (those that let a reader interact with components like graphics)~\citep{NolTem2007}. Because \cite{McN2016c} includes attributes `interactivity at every level', and the importance of `publishing' and `reproducibility', we would like tools to produce \emph{dynamic-interactive} graphics.


While we could imagine more graphical tools supporting reproducibility, the tools considered here are all formatting for files that can be compiled in some way to create a finished product.  

\subsection{knitr and RMarkdown}\label{rmarkdown}
The \proglang{R} community has long been committed to reproducibility, whether through simple script files containing analysis code or the \proglang{R} package \pkg{Sweave}~\citep{Lei2002}. \pkg{Sweave} allowed users to combine text and mathematical notation written in the markup language \LaTeX\, with \proglang{R} code into one source document. The source document could be processed to create a PDF document containing formatted text and math, \proglang{R} code, and output from \proglang{R} (for example, plots or numeric statistics). This system meant that the entire analysis writeup could be re-run to produce the same result. However, \pkg{Sweave} was fragile and difficult for novices to use. It has since been superseded by new developments. 

The \pkg{knitr} package by Yihui Xie has pushed the boundary even further~\citep{Xie2014}. \pkg{knitr} does everything \pkg{Sweave} did, but more generally and robustly. Where \pkg{Sweave} was limited to text written in \LaTeX\, and code written in \proglang{R}, \pkg{knitr} allows users to combine any type of code (\proglang{Python}, \proglang{C++}, etc) with any textual format. The most common textual formats are \LaTeX\, and Markdown. Markdown is much simpler than \LaTeX, so this makes \pkg{knitr} more accessible to novices. The most canonical examples of use are including \proglang{R} code in \LaTeX\, (the functionality that supersedes \pkg{Sweave}) or \proglang{R} code in \proglang{Markdown} text (called RMarkdown), but the package is much more flexible~\citep{Xie2014}.

Users write text and code (delimited as such by a particular syntax depending on the textual format they are using), then `knit' the source document to create a fully formatted \proglang{HTML}, PDF, or Word document. The output document has nicely formatted text, code with syntax highlighting, and all the results from the code, including numeric summaries and plots. If the user changes something in the source document, they have only to re-knit the document to see the updated results in their output document. \pkg{knitr} functionality is available through any \proglang{R} session, but the most embedded support is through RStudio

A specialized version of \proglang{Markdown} has been written to incorporate \proglang{R} code, called RMarkdown. Users can either `knit' a finished RMarkdown document to see their results, or execute code inline in a notebook setting much like the Jupyter notebooks in Section \ref{jupyter}. Introductory statistics students can use RMarkdown to submit their homework or produce reproducible reports for final projects~\citep{BauCet2014}. 

RMarkdown documents can also include interactive graphics produced using the \pkg{Shiny} package, discussed in more detail in Section \ref{shiny}. In this way, proficient \proglang{R} programmers can create dynamic-interactive documents. However, the code in knitted documents is static, so if readers want to interact with elements that were not programmed using \pkg{Shiny} they need the source file to be able to modify and re-knit or to use the notebook functionality to execute chunks inline. 

Whatever text markup language and programming language, \proglang{R} and \pkg{knitr} support reproducible research. \pkg{knitr} makes it simpler to share analysis results in such a way that the same analysis can be easily run on new data, changing only one line in the source code and re-knitting the report to see the results.

\subsection{Project Jupyter}
\label{jupyter}
Project Jupyter is focused on scientific computation and reproducibility~\citep{PerGra2015, RagPer2014}. It grew out of work in the \proglang{Python} community, and its best-known product is the Jupyter notebook (a next generation of the iPython notebook). Jupyter notebooks allow users to combine text and code in much the same way as the \pkg{knitr} package. The feature that separated \pkg{knitr} and Jupyter notebooks for several years was the fact that Jupyter notebooks were a single document, and cells of code could be executed to see results directly below the code~\citep{PerGra2007}. In contrast, \pkg{knitr} initially only supported the knitting of entire documents, not the execution of chunks of code within the source document. However, (perhaps inspired by Jupyter notebooks), RStudio now supports this type of notebook functionality in RMarkdown documents.

There are benefits and drawbacks to the notebook approach. The main advantage is that users can play with small pieces of code, and the manipulation is more direct because results are produced immediately below. This leads to the main weakness, because it is possible to execute code out of order and get to a state that would not be possible if the document was processed all the way through. This plays two of the attributes from \cite{McN2016c} off of one another. On one hand is interactivity, which notebooks certainly support. On the other is reproducibility, which the ability to play with code out of order can undermine. 

For the most robust support of Jupyter notebooks, users must install \proglang{Python} on their local machine. The Jupyter notebook is launched by typing \verb#jupyter notebook# in a terminal window. The installation and running of Jupyter therefore has more of a barrier to entry for novices than does \pkg{knitr} in RStudio. However, there are other ways to work in Jupyter. Instructors can host and support a server for students, or Project Jupyter hosts a sample version of the platform on the web, so anyone with internet access can try it. 

Jupyter notebooks initially only supported \proglang{Python} code, but are now more general. Official `kernels' are available for \proglang{Python}, \proglang{R}, and \proglang{Julia}, and community-maintained kernels support almost any other language you can think of (SAS software, \proglang{Sage}, \proglang{Go}, \proglang{Erlang}, and many more)~\citep{PerGra2015}. Users write text and code in `cells', which are flagged as Code, Markdown, Raw NBConvert, or Heading. In an interactive session, a user can execute code cells as they want to. To share their work, they can either provide the \verb#.ipynb# file to another user with Jupyter installed, or can export their work as HTML, PDF, or several other file formats. These exported documents are no longer interactive. 

Jupyter notebooks also provide the capability to create interactive graphics through the \pkg{IPython} library, so if the author has decided to include them readers can interact with selected graphics in the final product. Again, this may qualify them as dynamic-interactive documents. But, if a reader wants to interact with the code directly, they must get the \verb#.ipynb# file and work on it locally.  

\subsection{ODS, StatRep, StatWeave, MarkDoc}
\label{repro-s}
While \pkg{knitr}/RMarkdown and Jupyter notebooks support many programming languages, their integration with SAS software, Stata software, and SPSS has always been limited. However, there are other tools which have been developed to support reproducible research for those tools. 

The oldest feature from SAS software to support reproducible research was the development of the Output Delivery System (ODS). This allowed users of SAS software to print their results to many formats, including SAS data sets, \LaTeX, HTML, and RTF~\citep{BryMul2000}. Since then, the development of the StatRep package has enhanced reproducibility in SAS~\citep{ArnKuh2015}

In Stata software, users can create reproducible reports using StatWeave~\citep{Len2012}, which offers similar functionality to the \proglang{R} package \pkg{Sweave}. It allows users to combine \LaTeX\, and Stata code. However, StatWeave can be difficult to use, particularly for novices~\citep{Ris2014}. A newer development in the vein of \pkg{knitr} is MarkDoc, a way to combine Stata code with Markdown, HTML, or \LaTeX\, to create dynamic documents~\citep{Hag2016}.

All these tools are still under development, and are not as integrated into a graphical user interface or integrated development environment as RMarkdown or Jupyter notebooks. 

\subsection{Reflections on tools for reproducible research}
Current tools for reproducible research do a great job of combining text and code. The most common tools used by professional data analysts and statisticians are Jupyter notebooks and \pkg{knitr}/RMarkdown documents. There are other alternatives out there, like Beaker~\citep{TSOS2016} and Zeppelin~\citep{Zeppelin2016}.
More specific to SAS software and Stata software, packages like StatRep, StatWeave, and MarkDoc offer some of the same functionality, although not as fluidly. 

These tools all satisfy the `support for reproducibility' attribute from~\cite{McN2016c}. But, they all fail to varying degrees on `easy entry' and `interactivity at every level.' StatRep, StatWeave, and MarkDoc are all the hardest to use, because they do not have the interface supporting them that RMarkdown or Jupyter notebooks have. They require more steps to compile, and StatRep and StatWeave use \LaTeX\, (notoriously difficult to learn) as the text markup language.

RMarkdown documents, Jupyter notebooks, and their alternatives are easier for novices to use. They still require some coding finesse to use, but because the text markup language is markdown and the interface includes buttons to help add cells or code chunks, they can be used in introductory classes~\citep{BauCet2014}. 

All of these tools generally produce static documents, unless the author specifically codes in interactive features. Once a document is published or shared, the ability to execute code is removed, preventing readers from manipulating it. If a reader wants to modify the code, they must download the source code, edit it, and then re-share the results. So, they do not support dynamic-interactive documents~\citep{NolTem2007}. 

We are not aware of any graphical tools fully supporting reproducible research, although some of the bespoke tools mentioned in Section \ref{bespoke} support components of reproducible research graphically. 

\section{BESPOKE TOOLS}\label{bespoke}

In addition to the tools discussed above, there are a number of `bespoke' tools for doing particular things with data. These tools do not fall under any of the previous umbrellas, and represent some of the progress being made in statistical computing. Typically, these are standalone programs, mostly Graphical User Interfaces (GUIs), which are designed for doing one specific task along the data analytic pipeline. Although there are new bespoke tools popping up all the time, we will consider Data Wrangler, Open Refine, Tableau, and Lyra. 

\subsection{Data Wrangler/Trifacta}
Data Wrangler began as a project from the Stanford Visualization Group in 2011~\citep{KanPae2011}. Their goal was to provide a visual representation of data transforms, as well as a reproducible history of those transforms. For example, a user could select an empty row and indicate it should be deleted, at which point the Wrangler interface would suggest a variety of generalizable transformations that could be built from that one `rule' (e.g., delete all empty rows, or always delete the 7th row). Once the user specifies a transform, it is applied to the data and added to the interaction history. The interaction history can be exported as a data transformation script in a variety of languages. Wrangler can also perform simple database manipulations. 

The tools Wrangler provided were so useful the authors were able to convert their academic research project into a corporate venture, which is now known as Trifacta. Trifacta still offers a free version of their product, called Trifacta Wrangler, but their business model is building on selling their enterprise software. Pricing is not explicitly listed on their website, so companies interested in the product must contact the team to get a quote. Much like SAS software, Stata software, and SPSS, this model means that the free version could be used for teaching, but would essentially be grooming students to need an expensive product once they moved beyond the capabilities of the free version. 

Pricing notwithstanding, Wrangler provides the existence proof that a visual approach could be taken toward data cleaning while preserving reproducibility. 

\subsection{Open Refine}
Similar to Data Wrangler is Open Refine~\citep{VerDeW2013}. The project was initially called Google Refine, but has since been turned into an open source package. Like Wrangler, Open Refine can help clean data and document the data cleaning process. It can also be used for data exploration and data matching, including geooding. Again, the results of the refining process are available as a re-useable script. 

Both Data Wrangler and Open Refine provide great alternatives to the spreadsheet paradigm. They privilege data as a complete object, and document all modifications. By suggesting methods of generalizing data transformations, they remove much of the grunge work of spreadsheet analysis. The other benefit of generalized data transformations is they encourage the user to think computationally. Instead of just doing `whatever works,' there is user incentive to find a way to describe the data cleaning rule in a way that works generally. 

As its name suggests, Open Refine is open-source and free to use. However, since Open Refine (and, similarly, Wrangler) only does data transformation, using it in teaching would necessitate students learning an entire pipeline of products. Generally, we want to teach as few tools as possible, to reduce the overall cognitive load on students. No matter how good the second tool you teach to people is, they always like it less than the first one. 

\subsection{Tableau}
Tableau is a bespoke system for data visualization, based on~\cite{Wil2005}. As such, it does not provide much support for data cleaning. Tableau makes it simple for users to create interactive graphics that can be easily published on the web. It offers easy entry and simple support for publishing~\citep{McN2016c}. Tableau will suggest the `best' plot for particular data, which is both a blessing and a curse~\citep{MacHanSto2007}. It can lead to much more appropriate uses of standard plots, but it also does not support flexible plot creation. A user can make a plot without having any idea of what it means. Similarly, Tableau makes it possible to fit models to data, but again does not make it clear what these models mean or how appropriate they may be. Like other enterprise software, Tableau is expensive-- \$999 for an individual license or \$1,999 for an individual professional license. However, as with SAS software, they make the tool free to students. 

Tableau offers easy entry and interactivity, but is not as reproducible as the other tools mentioned in this section. It does make it possible to repeat the same analysis on another dataset by replacing the data source, but there is no way to audit the process otherwise. The processing that takes place is opaque, and can't be exported as human-readable code as in other systems. 

\subsection{Lyra}
\label{lyra}
Another bespoke data visualization system is Lyra, which makes it easy for novices to create graphics in a drag-and-drop matter~\citep{SatHee2014}. Lyra was developed at the University of Washington Interactive Data Lab. Interestingly, Jeffrey Heer was a member of the Stanford Visualization Group that created Data Wrangler, and is now one of the founders of Trifacta. He has since moved to the University of Washington and is a member of the Interactive Data Lab. 

Lyra is built on top of \proglang{vega}, an abstraction layer on top of \proglang{d3}, a \proglang{JavaScript} library. \proglang{d3} is a library for ``manipulating documents based on data,'' where `documents' refers to the document object model (DOM) of the web~\citep{BosOgi2011, Bos2013}. It is commonly used to create interactive web visualizations. \proglang{d3} is a very general library, and cannot be considered to be a plotting library at all. It does not provide primitives like bar, box, axes, etc., like standard visualization systems. Instead, it binds data to the DOM of a web page. 

Many of the interactive data visualizations by the New York Times (mentioned in Section \ref{interact}) are based on \proglang{d3}, and an online site allows  users can share `blocks' they have created in \proglang{d3}~\citep{Bos2015}. While the sharing of code examples helps users get started, \proglang{d3} is generally considered to be quite difficult to learn. 

\proglang{Vega} is an attempt to make it easier for novices to create the beautiful interactive graphics associated with \proglang{d3}~\citep{Hee2014}. It provides the sorts of graphical primitives more typically associated with data visualization tools: \verb#rect#, \verb#area#, and \verb#line#. However, even with these primitives, \proglang{Vega} can be difficult for novices in the same way all textual programming languages.

Enter Lyra, a tool to simplify the creation of \proglang{Vega} graphics. It supports simple data transformation, like grouping based on a variable, but generally should only be considered to be a visualization tool, because it does not provide functionality for data cleaning, modeling, etc. It is a reproducible tool, because the resulting graphics can be interrogated in the way standard \proglang{Vega} graphics can be (i.e., by looking at the code). Lyra does not support interactive graphics creation, but the group recently deployed an reactive version of \proglang{Vega}~\citep{SatRusHofHee2016, SatMor2016}, so it seems likely Lyra will soon go in that direction as well. 

Lyra provides much easier entry to making web graphics than tools like \proglang{d3}. It is close to being interactive at every level-- the process of creating visualizations is interactive, although the final product is not yet. Because the tool generates code as you move through the creation process, it is also reproducible.

\subsection{Reflections on bespoke tools}
Bespoke data tools like these are great sources for inspiration about new ways to visualize and improve data cleaning, modeling, and visualization. Many of these projects are open-source, and while they do not cover the entire analysis trajectory, they show promise as tools for particular data needs. We have focused here on Wrangler, Open Refine, Tableau, and Lyra, but there are many more bespoke projects out there. For example, \proglang{Brunel}~\citep{Brunel2016} is an alternative to \proglang{Vega} and Lyra. Like \proglang{vega}, \proglang{Brunel} offers a domain-specific language for visualization, and like Lyra, it also provides a graphic interface to the language. Another similar effort is \pkg{plot.ly}, a \proglang{JavaScript} plotting library that supports translation between graphics in \proglang{R}, \proglang{Python}, and \proglang{MATLAB}~\citep{Plotly2015}. As with Lyra and Brunel, \pkg{plot.ly} provides a graphical user interface to allow people not familiar with coding to create interactive graphics very simply. Again, code is generated (either in \pkg{plot.ly} syntax or a target language), so the process is reproducible. 

In fact, that is the inspiring element of many of the bespoke tools discussed here. They provide graphical user interfaces including new visual metaphors for data analysis alongside underlying code to provide reproducibility. They allow novices to perform complex data cleaning and visualization without getting lost in the syntactic weeds.

\section{CONCLUSIONS AND FURTHER WORK}

Given the attributes outlined in~\cite{McN2016c}, the existing tools used in statistics education once again break into two distinct groups-- tools for learning statistics, and tools for doing statistics. Tools that are interactive and offer easy entry are typically not flexible to extensions or reproducible. 

In particular, TinkerPlots and Fathom  forefront methods to increase the visual representation of analysis and to simplify it for novices. The bespoke tools Data Wrangler, Open Refine, Tableau, and Lyra also provide easy entry coupled with a more solid trace of the analysis. In contrast, more flexible, scriptable tools like \proglang{R}, SAS software, Stata software, or SPSS are harder to get started using and much less interactive. 

The bespoke products we examined here provide inspiration that a tool could satisfy all 10 of the attributes at once (perhaps with varying levels of success). I want to encourage statistics educators to look to the future and consider what an ideal tool might look like, several years down the line. One vision would be a blocks-based language providing drag-and-drop functionality for novices, with a domain-specific language underlying it for more advanced students leading to a target language used by professionals. More of this vision is shared in~\cite{McN2016c}. 

However, since we all live in the present, it seems important to offer best practices given current computational tools. 

As is probably clear, my preference for statistics education is \proglang{R}, using the formula syntax or the piped tidyverse syntax. Whichever syntax is chosen, educators should make every attempt to only expose students to that one syntax. In an introductory course, this is possible. Instructors using the OpenIntro textbook Introduction to Statistics with Randomization and Simulation have written the associated labs in a variety of `flavors' to limit students' exposure to a particular syntax~\citep{DieBar2014}. For interactive work, \pkg{Shiny} or \pkg{manipulate} can be used for applet-like functionality. Best practices also include the use of the IDE for \proglang{R}, RStudio, and having students produce reproducible work using RMarkdown~\citep{BauCet2014}. 


For instructors who are not willing to make the leap to a programming language, I believe the best existing tool to use at the college level is Fathom (TinkerPlots is very similar and could be substituted, but is aimed at slightly younger students). Fathom offers easy entry, lots of visual cues, and encourages iteration and randomization. It provides flexible and creative ways to explore data. The lack of support for reproducible analysis or the sharing of results is problematic, but for students not continuing on in statistics, this may be acceptable. 

Not recommended are applets (other than as demos by an instructor), graphing calculators, or spreadsheet software.


\bibliography {../../../ReadingLibrary}

\end{document}